\documentstyle[prd,aps]{revtex}

\newcommand{\bea}{\begin{eqnarray}}
\newcommand{\eea}{\end{eqnarray}}

\begin{document}
\draft
\twocolumn[\hsize\textwidth\columnwidth\hsize\csname
@twocolumnfalse\endcsname

\title{Cosmological structure problem of the ekpyrotic scenario}
\author{Jai-chan Hwang${}^{(a,b)}$ \\
        ${}^{(a)}$ Department of Astronomy and Atmospheric Sciences,
                   Kyungpook National University, Taegu, Korea \\
        ${}^{(b)}$ Institute of Astronomy, Madingley Road, Cambridge, UK}
\date{\today}
\maketitle

\begin{abstract}

We address the perturbation power spectrum generated in the
recently proposed ekpyrotic scenario by Khoury {\it et al}. The
issue has been raised recently by Lyth who used the conventional
method based on a conserved variable in the large-scale limit, and
derived different results from Khoury {\it et al}. The calculation
is straightforward in the uniform-curvature gauge where the
generated blue spectrum with suppressed amplitude survives as the
final spectrum. Whereas, although the metric fluctuations become
unimportant and a scale-invariant spectrum is generated in the
zero-shear gauge the mode does not survive the bounce, thus with
the same final result. Therefore, an exponential potential leads
to a power-law expansion/contraction $a \propto |t|^p$, and the
power $p$ dictates the final power spectra of both the scalar and
tensor structures. If $p \ll 1$ as one realization of the
ekpyrotic scenario suggests, the results are $n_S -1 \simeq 2
\simeq n_T$ and the amplitude of the scalar perturbation is
suppressed relative to the one of the gravitational wave by a
factor $\sqrt{p}/2$. Both results confirm Lyth's. An observation
is made on the constraint on the dynamics of the seed generating
stage from the requirement of scale-invariant spectrum.

\end{abstract}

\vskip2pc]

\section{Introduction}

The ekpyrotic universe scenario based on colliding branes imbedded
in extra-dimensional bulk has been suggested recently by Khoury
{\it et al.} in \cite{KOST}. Perhaps because of its ambitious plan
to explain the origin of the hot big bang, and also of its plan to
generate the scale-invariant (Harrison-Zeldovich \cite{Harrison-Zeldovich})
spectrum without resorting to the inflation-type accelerating stage
\cite{KOST-pert}, it has been under close examinations
\cite{others}. In particular, a quite different scalar spectrum,
including both the amplitude and the slope, was derived by Lyth in
\cite{Lyth-1} which is supposed to be fatally threatening the
scenario as a viable addition to the early universe models, see
also \cite{BF,Lyth-2} for recent additional arguments against
\cite{KOST,KOST-pert}.

If the imbedded 3 space literally passes through the singularity
during the collision phase of the branes, thus making the 3+1
dimensional equations obsolete, probably we do not have handle
about how to calculate the generated spectra from the scenario. It
is suggested that the ekpyrotic scenarios
\cite{KOST,KOST-pert,KOSST} in fact go through singularity in the
view point of our three space, and in such a case we anticipate
serious problems associated with the breakdown of the basic
equations we are using. In this sense, to address the structural
seed generation mechanism properly in the ekpyrotic scenario, it
is likely that we need to handle the perturbation analyses in the
context of the higher dimension which is at the moment an
unsettled issue. 
If we {\it assume}, however, that the 3+1
dimensional effective field theoretical description works, and the
linear perturbation theory holds during the bouncing stage, we can
apply some well known tools of the cosmological perturbations
developed especially over the last two decades. The current
controversy about the scalar spectrum is in this narrow context
\cite{KOST,KOST-pert,Lyth-1,BF} which we will also accept in the
following. The possibility of breakdown of the linear perturbation
theory as the scale factor approaches singularity was pointed out
by Lyth in \cite{Lyth-2}. 
If we agree, however, that we can handle
the situation using the linear perturbation 
based on the 3+1 spacetime effective theory with
nonsingular bounce, as we will show below, the results are already
well known in the literature which are used to make correct
estimations \cite{Lyth-1,BF}.

An exponential type potential leads to a power-law expansion
\cite{Lucchin-Matarrese-1985}
\bea
   & & a \propto |t|^p, \quad
       V = - {p ( 1 - 3 p) \over 8 \pi G} e^{-\sqrt{16 \pi G /p} \phi},
   \label{power-law}
\eea
which includes the contraction as well.
With $p \ll 1$ this potential is an example of the ekpyrotic
scenario considered in \cite{KOST,KOST-pert}.
Assuming (i) both the
scalar and the tensor perturbations were generated from quantum
fluctuations (of the field and the metric) during such a power-law
era, and were pushed outside horizon, the analytic forms of the
spectra based on the vacuum expectation values are known in the
literature, see eqs. (47,48) in \cite{Stewart-Lyth-1993} for a summary.
Under the simplest vacuum state, we have
\bea
   {\cal P}_{\hat \varphi_v}^{1/2}
   &=& \sqrt{4 \pi G p} {|H| \over 2 \pi}
       2^{-1 \over 1 - p} \left| {1 - p \over p} \right|^{-p \over 1 - p}
       {\Gamma ({1 - 3p \over 2(1 - p)} ) \over \Gamma(3/2)}
   \nonumber \\
   & & \times
       \left[ {k / (a|H|)} \right]^{1 \over 1- p},
   \label{P_S} \\
   {\cal P}_{\hat C_{\alpha\beta}}^{1/2} 
   &=& ({2 / \sqrt{p}}) {\cal P}_{\hat \varphi_v}^{1/2},
   \label{P_T}
\eea
where $C_{\alpha\beta}$ is the tensor-type metric fluctuations,
and $\varphi_v \equiv \varphi - (aH/k) v$ introduced by Lukash in
\cite{Lukash-1980} is a gauge-invariant combination which is
proportional to the perturbed three-space curvature ($\varphi$) in
the comoving gauge ($v \equiv 0$); $\varphi_v$ is also introduced
as $\phi_m$ in \cite{Bardeen-1980}, and is the same as ${\cal R}$
in \cite{Stewart-Lyth-1993}. 
{}For our notation, see \cite{Noh-Hwang-2001-Unified}.
Overhats indicate the spectra based
on the vacuum expectation value of the quantum fluctuations of the
field and the metric. Now, assuming (ii) the scale stays outside
horizon while there are transitions (like the inflation to the
radiation dominated eras, and radiation to the matter dominated
eras), due to the conservation property of the growing solutions of
both $\varphi_v$ and $C_{\alpha\beta}$ [see eqs.
(\ref{LS-sol-varphi_v},\ref{GW-sol})] we can identify the above
spectra ${\cal P}_{\hat \varphi_v}$ and ${\cal P}_{\hat
C_{\alpha\beta}}$ imprinted during the quantum generation stage
just after the horizon crossing with the classical power spectra
${\cal P}_{\varphi_v}$ and ${\cal P}_{C_{\alpha\beta}}$ based on
on the spatial averages at the second horizon crossing epoch.
Therefore, spectral indices for the scalar ($S$) and the tensor
($T$) structures, $n_S - 1 \equiv d \ln{{\cal P}_{\varphi_v}} /d
\ln{k}$ and $n_T \equiv d \ln{{\cal P}_{C_{\alpha\beta}}} /d
\ln{k}$, become
\bea
   & & n_S - 1 = {2 /(1 - p)} = n_T.
   \label{spectra}
\eea
Thus, in the power-law inflation limit with large $p$ we have the
scale-invariant spectra $n_S - 1 \simeq 0 \simeq n_T$.

In the ekpyrotic scenario, although there is no acceleration phase
before the radiation dominated big bang stage, the two assumptions
(i,ii) above apply as well; the growing solutions of $\varphi_v$
and $C_{\alpha\beta}$ are conserved as long as we have the
large-scale conditions (see later) are met, and we will see that
these conditions are satisfied during the transition from the
collapsing to the expanding phases in the ekpyrotic scenario.
Thus, now with $p \ll 1$ we have $n_S - 1 \simeq 2 \simeq n_T$
which differs from the result in \cite{KOST,KOST-pert} for $n_S$.
Besides the wrong spectral slope, eq. (\ref{P_T}) shows that the
amplitude of scalar structure is suppressed relative to the one of
gravitational wave, which probably means that the scalar
perturbation should be negligible as pointed out in \cite{Lyth-1}.

We still observe the different opinions maintained in the
literature: (i) the final scale-invariant spectrum from the
scenario \cite{KOST,KOST-pert}, (ii) the final blue spectrum with
negligible amplitude when the bounce is nonsingular
\cite{Lyth-1,BF}, and (iii) the breakdown of the linear theory in
the singular bounce \cite{Lyth-2}. The issues are manifold
involving different gauge conditions, different matching
conditions used, and others. In the following we will address the
issues involved in the differences and will indicate how the
analyses consistently support (ii) or (iii) instead of (i). Since
there is no controversy over the tensor spectrum, we will
concentrate on the scalar spectrum.

\section{Quantum generation: two gauges}
                                                    \label{sec:Quantum}

The different opinions between \cite{KOST,KOST-pert} and
\cite{Lyth-1,BF} can be partly traced to different generated
spectra before the model makes the bounce. In a naive calculation
ignoring the metric fluctuations, \cite{KOST} showed a
scale-invariant spectrum. In a rigorous calculation (based on
$\delta \phi$ in the uniform-curvature gauge), \cite{Lyth-1},
however, showed that including the metric fluctuation is important
resulting in a blue spectrum with suppressed amplitude. Then,
\cite{KOST-pert} showed that, in fact, in the zero-shear gauge the
metric fluctuation becomes negligible, thus confirming the
original scale-invariant spectrum; see below. 
Different results from different gauge conditions in
the large-scale limit are not surprising because the metric
fluctuations often dominate in that scale. Although the final
observable result should be the same, the intermediate steps could
depend on the gauge conditions we choose for the analyses. In the
case of the uniform-curvature gauge the later evolution is simple
as explained below eq. (\ref{P_T}), and the generated spectrum
simply survives as the final spectrum, whereas the analyses in the
zero-shear gauge is somewhat intricated which we will explain in
the following and the next section.

In order to clarify the situation, in this section we present the
generated scalar perturbation during the quantum generation stage
in the two gauge conditions. 
We introduce
\bea
   & & \varphi_{\delta \phi} \equiv \varphi
       - ({H / \dot \phi}) \delta \phi
       \equiv - ({H / \dot \phi}) \delta \phi_\varphi,
   \label{GI-def}
\eea
which are just different definitions of a gauge-invariant
combination of $\varphi$ and $\delta \phi$; $\varphi_{\delta
\phi}$ can be interpreted as the $\varphi$ in the uniform-field
gauge ($\delta \phi = 0$), and $\delta \phi_\varphi$ is the
$\delta \phi$ in the uniform-curvature gauge ($\varphi = 0$); as
$v = (k/a) \delta \phi/\dot \phi$ for the field, we have
$\varphi_{\delta \phi} = \varphi_v$. The above relation is
powerful in analyzing the classical evolution and the quantum
generation of scalar perturbation, see
\cite{Noh-Hwang-2001-Unified} for a summary. Whereas, as we will
explain later, using $\delta \phi_\chi \equiv \delta \phi - \dot
\phi \chi$, which is $\delta \phi$ in the zero-shear gauge ($\chi
\equiv 0$), the analysis becomes somewhat involved eventually
leading to the same final result.

The equation for the perturbed scalar field is simplest when viewed
in the uniform-curvature gauge \cite{Hwang-1993-QFT}
\bea
   & & \delta \ddot \phi_\varphi + 3 H \delta \dot \phi_\varphi
       + \left( {k^2 / a^2} + V_{,\phi\phi} \right) \delta \phi_\varphi
   \nonumber \\
   & & \qquad
       + 2 {\dot H \over H} \left( 3 H - {\dot H \over H}
       + 2 {\ddot \phi \over \dot \phi} \right) \delta \phi_\varphi = 0,
   \label{delta-phi-eq-UCG}
\eea
where the terms in the second line come from the metric
perturbations, compare eqs. (7,A9) in \cite{Hwang-1993-QFT}; other
gauge conditions cause more complicated contributions from the
metric \cite{Hwang-1994-MSF}. Calling the metric term a metric
back-reaction could be misleading because the perturbed field
excites/accompanies the metric fluctuations simultaneously.
Keeping the contribution from the metric is necessary and makes
the equation consistent and even simpler in the sense that we have
a general large-scale solution, see eq. (\ref{LS-sol-varphi_v}).
When the background is supported by a near
exponential expansion 
the whole term from the metric, and $V_{,\phi\phi}$ separately,
nearly vanish; this explains why the original derivation of the
inflationary spectra in \cite{inflation-pert,BST-1983} was
successful even without fully considering the metric perturbations. 
However, situation could be different in other
cases like the power-law expansion (contraction as well) where the
ekpyrotic scenario based on exponential potential is one example.
In the power-law expansion in eq. (\ref{power-law}) the metric
term {\it cancels} with $V_{,\phi\phi}$ exactly, see eq. (22) in
\cite{Hwang-1993-QFT}. It happens that for $p \ll 1$ the
$V_{,\phi\phi}$ term without the metric term gives a contribution
which can be translated to the $n_S \simeq 1$ scalar power
spectrum; this was pointed out in \cite{Lyth-1}. 
However, this term should be cancelled exactly by the metric term. 
With this metric effect taken into account we end up with a massless free
scalar field equation which can be translated to $n_S \simeq 3$
generated spectrum.
As explained below eq. (\ref{P_T}) this generated spectrum 
simply survives the later evolution, see next section.

In the zero-shear gauge the equation for $\delta \phi_\chi$ was
derived in eq. (27) of \cite{Hwang-1993-QFT}
\bea
   & & \delta \ddot \phi_\chi + \left[ 3 H
       + {8 \pi G (\ddot \phi + H \dot \phi) \over \dot H + k^2/a^2}
       \dot \phi \right] \delta \dot \phi_\chi
   \nonumber \\
   & & \quad
       + \left[ {k^2 \over a^2} + V_{,\phi\phi} + 4 \dot H
       - {8 \pi G (\ddot \phi + H \dot \phi) \over \dot H + k^2/a^2}
       \ddot \phi \right] \delta \phi_\chi = 0,
   \label{delta-phi-eq-ZSG}
\eea
which looks quite complicated compared with eq. (\ref{delta-phi-eq-UCG}).
In the small-scale limit, eqs. (\ref{delta-phi-eq-UCG},\ref{delta-phi-eq-ZSG})
both reduce to the massless and free scalar field equation.
In the large-scale limit, we have
\bea
   & & \delta \ddot \phi_\chi + 3 H \delta \dot \phi_\chi
       + V_{,\phi\phi} \delta \phi_\chi + 4 \dot H \delta \phi_\chi = 0.
   \label{delta-phi-eq-ZSG-LS}
\eea
Thus, in this limit, the only metric contribution which is the
last term becomes simple. As we have $V_{,\phi\phi} = -2 (1 -
3p)/t^2$ and $4 \dot H = - p/t^2$, for $p \simeq 0$ the metric
contribution becomes negligible compared with $V_{,\phi\phi}$;
this is in contrast with the uniform-curvature gauge case where
the metric term cancels exactly with the $V_{,\phi\phi}$ term. The
authors of \cite{KOST-pert} pointed out that the analysis in this
gauge, which is rigorous, is the same as the one based on naive
calculation simply ignoring the metric perturbations presented in
\cite{KOST}. Thus, the generated spectrum for $\delta \phi_\chi$
is scale-invariant, whereas the one for $\delta \phi_\varphi$ is
blue. Later, we will show how these two apparently different
results lead to the same final observable spectrum in the
expanding phase.

Now, we present a rigorous derivation of the generated perturbation
in the two gauges.
Using Mukhanov's notation in \cite{Mukhanov-1988} we have, see also
eqs. (46,47,68-70) in \cite{Hwang-1994-MSF},
\bea
   & & u = - {4 \pi G z \over k^2} \left({v \over z} \right)^\prime,
       \quad
       v = {1 \over 4 \pi G z} \left( z u \right)^\prime,
   \label{u-v-relations} \\
   & & v^{\prime\prime} + \left[ k^2 - {z^{\prime\prime} \over z}
       \right] v = 0, \;\;
       u^{\prime\prime} + \left[ k^2 - {(1/z)^{\prime\prime} \over 1/z}
       \right] u = 0,
   \label{u-v-equations} \\
   & & v \equiv a \delta \phi_\varphi, \quad
       u \equiv - \varphi_\chi/\dot \phi, \quad
       z \equiv {a \dot \phi / H},
   \label{u-v-definitions}
\eea
where $\varphi_\chi \equiv \varphi - H \chi$, and a prime
indicates a time derivative based on $\eta$ where $dt \equiv a d \eta$. 
We have
\bea
   & & \delta \phi_\chi = \delta \phi_\varphi + (\dot \phi / H) \varphi_\chi,
   \label{delta-phi_chi}
\eea
which follows from eq. (\ref{GI-def}) evaluated in the zero-shear gauge.
In the power-law case, using eqs. (7,8) of \cite{KOST-pert}
we have ($p \neq 1$)
\bea
   & & {z^{\prime\prime} \over z} = - {p(1 - 2p) \over (1 - p)^2}
       {1 \over \eta^2}, \quad
       {(1/z)^{\prime\prime} \over 1/z} = {p \over (1 - p)^2}
       {1 \over \eta^2}.
   \label{z-power-law}
\eea
Notice that eqs. (\ref{u-v-equations},\ref{z-power-law}) lead to
Bessel equations for $v$ and $u$ with different orders. Using
the quantization based on the action formulation of Mukhanov
in \cite{Mukhanov-1988}, we have the mode function
solutions
\bea
   \delta \phi_{\varphi k} 
   &=& {\sqrt{ \pi |\eta|} \over 2 a}
       \left[ c_1 (k) H_{\nu_v}^{(1)} (k|\eta|)
       + c_2 (k) H_{\nu_v}^{(2)} (k|\eta|) \right],
   \nonumber \\
   \varphi_{\chi k} 
   &=& {|H| \sqrt{\pi|\eta|} \over 2 k \sqrt{2p} M_{pl}}
       \Big[ c_1 (k) H_{\nu_u}^{(1)} (k|\eta|)
       + c_2 (k) H_{\nu_u}^{(2)} (k|\eta|) \Big],
   \nonumber \\
   & & \nu_v = {3 p - 1 \over 2 (p - 1)}, \quad
       \nu_u = {p + 1 \over 2 (p - 1)}.
   \label{solutions}
\eea
The quantization condition implies
$|c_2|^2 - |c_1|^2 = \pm 1$ where the sign corresponds to
the sign of $\eta$;
the positive frequency Minkowski space mode in the small-scale
limit corresponds to $c_1 = 0$ for positive $\eta$ and
$c_2 = 0$ for negative $\eta$.
The other variables $\varphi_{\delta \phi}$ and
$\delta \phi_\chi$ can be recovered using 
eqs. (\ref{GI-def},\ref{delta-phi_chi}).
One can check that the solutions in eqs. (\ref{solutions},\ref{delta-phi_chi}) 
are consistent with eq. (\ref{u-v-relations}).

The power-spectrum from vacuum
quantum fluctuation is related to the mode function solution as
${\cal P}_{\delta \hat \phi_\varphi} = {k^3/(2 \pi^2)}
|\delta \phi_{\varphi k}|^2$ and similarly for
${\cal P}_{\delta \phi_\chi}$, \cite{Hwang-1993-QFT}.
Thus, in the large-scale limit we can derive
(for $\nu = 0$ we have an additional $2 \ln(k|\eta|)$ factor)
\bea
   {\cal P}_{\delta \hat \phi_\varphi}^{1/2}
   &=& {|H| \over 2 \pi} \left| {1 - p \over p} \right|
       {\Gamma (\nu_v) \over \Gamma (3/2)}
       \left( {k |\eta| \over 2} \right)^{1 \over 1 - p}
   \nonumber \\
   & & \times |c_2 - c_1|, \quad \nu_v \ge 0,
   \label{P_UCG1} \\
   &=& {|H| \over 2 \pi} \left| {1 - p \over p} \right|
       {\Gamma (-\nu_v) \over \Gamma (3/2)}
       \left( {k |\eta| \over 2} \right)^{2 - 3p \over 1 - p}
   \nonumber \\
   & & \times |c_2 e^{i \nu_v \pi} - c_1 e^{-i \nu_v \pi}|, \quad \nu_v \le 0,
   \label{P_UCG2} \\
   {\cal P}_{\delta \hat \phi_\chi}^{1/2}
   &=& {|H| \over 2 \pi} {1 \over 2}
       {\Gamma (\nu_u) \over \Gamma (3/2)}
       \left( {k |\eta| \over 2} \right)^{1 \over 1 - p}
   \nonumber \\
   & & \times |c_2 - c_1|, \quad \nu_u \ge 0,
   \label{P_ZSG1} \\
   &=& {|H| \over 2 \pi} {1 \over 2 |p|}
       {\Gamma (-\nu_u) \over \Gamma (3/2)}
       \left( {k |\eta| \over 2} \right)^{-p \over 1 - p}
   \nonumber \\
   & & \times |c_2 e^{i \nu_u \pi} - c_1 e^{-i \nu_u \pi}|, \quad \nu_u \le 0.
   \label{P_ZSG2}
\eea
The spectral index of the generated fluctuation is $n_{\delta
\phi_\varphi} - 1 = d \ln{ {\cal P}_{\delta \hat \phi_\varphi} }
 /( d \ln k )$ and similarly for $n_{\delta \phi_\chi}$.
By taking the vacuum state to match with the positive frequency
Minkowski space mode, thus $|c_2 - c_1| = 1$, etc., eqs.
(\ref{P_UCG1},\ref{GI-def}) give results in eqs.
(\ref{P_S},\ref{spectra}), whereas eq. (\ref{P_ZSG2}) is the one
derived in eq. (17) of \cite{KOST-pert} for $p \simeq 0$. Both
results are consistent, except that what we need for later
surviving spectrum is the one for $\delta \phi_\varphi$, which is
directly related to $\varphi_{\delta \phi}$, and not the one for
$\delta \phi_\chi$ which does not survive, see next section. That
is, for $p \simeq 0$, as we follow the later evolution we can show
that $n_S = n_{\varphi_{\delta \phi}} = n_{\delta \phi_\varphi}
 \neq n_{\delta \phi_\chi}$.
One important fact to notice is that the spectra in 
eqs. (\ref{P_UCG1},\ref{P_ZSG1}) are time independent,
whereas the ones in eqs. (\ref{P_UCG2},\ref{P_ZSG2}) are time dependent.

\section{Classical evolution and final spectrum}

Since the final scalar spectrum could be directly related to the large
angular scale CMB anisotropies the final result should be physical,
and the methods (involving the gauges, matching conditions, etc.)
used to get the results should not affect the final results.
In this section, assuming the linear perturbation theory is
valid for scales we are interested in, we present the evolution of the
perturbation generated during the collapsing phase
as the background model goes through (smooth and nonsingular) bounce
into the expanding phase.

In the large-scale limit, thus ignoring $k^2$ terms, eq.
(\ref{u-v-equations}) have general solutions
\cite{Noh-Hwang-2001-Unified}
\bea
   \varphi_\chi (k, t)
   &=& {4 \pi G \mu a^2 \over k^2 - 3 K} \delta_v (k, t)
   \nonumber \\
   &=& 4 \pi G C (k) {H \over a} \int^t {a(\mu + P) \over H^2} dt
       + {H \over a} d (k),
   \label{LS-sol-varphi_chi} \\
   \varphi_v (k, t) 
   &=& C (k) - d(k) {k^2 \over 4 \pi G} \int^t {d t \over a^3 Q},
   \label{LS-sol-varphi_v} \\
   \varphi_\delta (k, t)
   &=& \varphi_v + {1 \over 12 \pi G (\mu + P)} {k^2 \over a^2} \varphi_\chi,
   \label{LS-sol-varphi_delta}
\eea
where $Q = {\mu + P \over c_s^2 H^2}$ with $c_s^2 \equiv \dot
P/\dot \mu$ for the fluid, and $Q = {\dot \phi^2 /H^2}$ for the
field; for notations, see\footnote{
         $\delta_v \equiv \delta + 3 (aH/k) (1 + w) v$ is the same as 
         the density perturbation ($\delta$) in the comoving gauge; 
         $w \equiv P/\mu$, $P$ the pressure and $\mu$ the energy density,
         $\delta \equiv \delta \mu/\mu$.
         $\varphi_\delta \equiv \varphi + \delta/[3(1+w)]$ is the same as
         $\varphi$ in the uniform-density gauge ($\delta \equiv 0$).
         In Bardeen's notation \cite{Bardeen-1980} we have
         $\varphi_\chi = \Phi_H$ and $\delta_v \equiv \epsilon_m$.
         $\varphi_\delta$ is the same as $\zeta$ introduced by Bardeen in
         \cite{Bardeen-new}, and for $K = 0 = \Lambda$ it becomes the one
         introduced in \cite{BST-1983}.
         Notice that $\zeta$ used in \cite{KOST-pert,BF} is our $- \varphi_v$.
      }.
$C$ and $d$ are spatially dependent two integration constants. We
call the term with coefficient $C$ the $C$-mode and the other the
$d$-mode.  The $C$-mode is relatively growing and the $d$-mode is
decaying in an expanding phase, whereas the $C$-mode is
relatively decaying and the $d$-mode can grow in a contracting phase. 
We emphasize the general character of these solutions
which are valid considering generally time varying equation of
state $P(\mu)$ or potential $V(\phi)$. 
Above results are valid for $K = 0$; for more general forms applicable to
general $K$ and $\Lambda$, see \cite{Noh-Hwang-2001-Unified}. 
Notice that the $d$-modes of
$\varphi_v$ and $\varphi_\delta$ are already higher order in the
large-scale expansion compared with the one of $\varphi_\chi$.

As the large-scale power spectra in eqs.
(\ref{P_UCG1},\ref{P_ZSG1}) are time-independent, these can be
identified with the constant $C$-modes of $\varphi_v$ and
$\varphi_\chi$ in eqs.
(\ref{LS-sol-varphi_v},\ref{LS-sol-varphi_chi}); in our power-law
background we have $\delta \phi_\varphi \propto \varphi_v$ and
$\delta \phi_\chi \propto \varphi_\chi$ which follow from eqs.
(\ref{GI-def},\ref{delta-phi_chi}). 
Thus, for $p \gg 1$, eqs.
(\ref{P_UCG1},\ref{P_ZSG1}) both give the same final
scale-invariant spectra in eq. (\ref{spectra}). 
Whereas, notice the time dependences
of eqs. (\ref{P_UCG2},\ref{P_ZSG2}) which are proportional to $|t|/a^3$
and $|H|/a$.
These can be identified with the $d$-modes of
$\varphi_v$ and $\varphi_\chi$ in eqs.
(\ref{LS-sol-varphi_v},\ref{LS-sol-varphi_chi}); the latter one
was correctly pointed out in \cite{BF}, and a scale-invariant case
with $p = {2 \over 3}$ for $\varphi_v$ corresponds to the
$d$-mode, thus not interesting. 
{}For the ekpyrotic scenario with
$p \ll 1$ we have eq. (\ref{P_UCG1}) for $\delta \phi_\varphi$
thus $\varphi_v$, and eq. (\ref{P_ZSG2}) for $\delta \phi_\chi$
thus $\varphi_\chi$.  
Hence, while the power spectrum for
$\varphi_v$ contributes to the $C$-mode, the one for
$\varphi_\chi$ contributes to the $d$-mode.  
As will be further
explained later, despite its apparent growth in time the $d$-mode
generated in a contracting phase is not interesting because it
will affect only the $d$-mode which is a real decayng mode in the
later expanding phase.

In the case of $p<1$, since ${k \over a|H|} \propto |\eta| \propto |t|^{1-p}$
becomes small\footnote{
        In order to tell the large-scale nature,
        what we have to compare is the $z^{\prime\prime}/z$ and
        $z(1/z)^{\prime\prime}$ terms
        relative to $k^2$ term in eq. (\ref{u-v-equations}).
        }
as we approach the bouncing epoch, the large-scale general
solutions in eqs.
(\ref{LS-sol-varphi_chi}-\ref{LS-sol-varphi_delta}) are well valid
considering time-varying $P(\mu)$ or $V(\phi)$ including sudden
jumps. Since the solution is valid considering general
time-varying equation of state or potential, the vanishing
potential near bounce in \cite{KOST-pert} will not affect the
final result as long as the scale remains in the large-scale.

As we already have the general solutions the matching
approximation is unnecessary for the case which is supposed to be
an approximation. 
However, since \cite{KOST-pert} employed some 
{\it ad hoc} matching conditions to make the scale-invariant spectrum of
$\varphi_\chi$ to survive as the dominant mode in the expanding
phase, in the following we will explain how the {\it proper}
matching conditions lead to a consistent result with the one based
on the general solutions in eqs.
(\ref{LS-sol-varphi_chi}-\ref{LS-sol-varphi_delta}).

In \cite{Hwang-Vishniac-1991}
two gauge-invariant joining variables were derived which
are continuous at the transition accompanying a discontinuous
change in pressure assuming perfect fluids.
These are
\bea
   \varphi_\chi, \quad \varphi_\delta.
   \label{joining}
\eea
These are shown to be continuous for general $K$ and $\Lambda$ in arbitrary
scale. 
Instead of $\varphi_\chi$ we can use $\delta_v$ in 
eq. (\ref{LS-sol-varphi_chi}) as the continuous variable as well. 
The transitions between scalar
fields, and between the fluid and the field are treated
separately, see below eq. (15) of \cite{Hwang-Vishniac-1991}.
{}For the background, $a$ and $\dot a$ should be continuous at the transition. 
Consider two phases $I$ and $II$ with different
equation of states, making a transition at $t_1$. In the
large-scale limit by matching $\varphi_\chi$ and $\varphi_\delta$
in eqs. (\ref{LS-sol-varphi_chi}-\ref{LS-sol-varphi_delta}) we can
see that to the leading order in the large-scale expansion we have
\bea
   C_{II} 
   &=& C_{I}
   \nonumber \\
   d_{II} 
   &=& d_{I} + 4 \pi G C_{I} \Bigg[
       \int^{t_1} {a (\mu + P) \over H^2} dt \Bigg|_{I}
   \nonumber \\
   & & - \int^{t_1} {a (\mu + P) \over H^2} dt \Bigg|_{II} \Bigg].
   \label{matching-1}
\eea
This is consistent with the result in eq. (17) of
\cite{Hwang-Vishniac-1991}.  Thus, to the leading order in the
large-scale expansion the $C$-mode of $\varphi_v$ remains the
same, whereas the $d$-mode of $\varphi_\chi$ is affected by the
transition and also the previous history of the $d$- and $C$-modes
\cite{Hwang-Vishniac-1991}. 
Thus, the evolution of the $C$-mode
does not depend on the intermediate stages while the perturbations
are in the large-scale.  
While in the super-horizon scale the
effect of the entropic term of the scalar field is negligible,
thus the same conclusions apply to the case including the field as
well.

These results from the joining method coincide with the general
large-scale conservation solutions in eqs.
(\ref{LS-sol-varphi_chi},\ref{LS-sol-varphi_v},\ref{LS-sol-varphi_delta})
which are valid for the time varying equation of state $P(\mu)$ or
potential $V(\phi)$. Thus, our joining method simply confirms that
by using the {\it proper} joining variables we can recover the
correct results. We note that the results based on the integral
solutions or the joining methods are {\it not sensitive} to whether
the background is expanding or collapsing. Only condition
required is the large-scale condition where we could ignore the
$k^2$-term in the perturbation equation. Analyses made in
\cite{BF} confirm our general conclusions above in the specific
situation of the ekpyrotic scenario.

We note that in the collapsing background, the $d$-modes in eq.
(\ref{LS-sol-varphi_chi}) grows in time; it is called the growing
solution in \cite{KOST-pert,BF}. We have shown that the leading
order scale-invariant spectrum of $\delta \phi_\chi$ generated in
the collapsing phase with $p \ll 1$ should be identified as the
$d$-mode.  Despite its apparent growth in time we are not
interested in this $d$-mode based on the following reasons.
{}Firstly, eq. (\ref{LS-sol-varphi_chi}) is a general solution
valid for time-varying equation of state or potential, thus
independently of whether the time-dependence is growing or
decaying the $d$-mode remains as the $d$-mode which decays in
the eventual expanding phase. 
Secondly, by using the sudden jump
approximation we have shown that the $d$-mode does {\it not}
influence the $C$-mode which is the proper relatively growing mode
in expanding phase. Thus, $d$-mode is uninteresting afterall; it
dies away in a few Hubble expansion as the model enters the
expanding phase.

In \cite{KOST-pert} the authors {\it proposed} to use the
non-divergent variables at the transition as the joining variable,
and arrived at matching the two coefficients of a variable
$\delta_v$, see below eq. (43) of \cite{KOST-pert}. In a single
component situation we need two matching conditions, and we should
use the matching conditions on two independent variables, not on
the two coefficients of one variable. The matching conditions {\it
proposed} in \cite{KOST-pert} are {\it different} from the ones in
eq. (\ref{joining}) and we doubt their validity. 
Near bounce of
ekpyrotic scenario, as the potential nearly vanishes (see
\cite{KOST-pert}) we have $p \simeq {1 \over3}$ and the $d$-modes
of $\varphi_\chi$, $\varphi_v$ and $\varphi_\delta$ in eqs.
(\ref{LS-sol-varphi_chi}-\ref{LS-sol-varphi_v}) diverge whereas
$\delta_v$ is finite due to multiplication of a vanishing factor
$1/(\mu a^2)$ in eq. (\ref{LS-sol-varphi_chi}).
In passing we note that the corresponding solution of the
gravitational wave is \cite{Noh-Hwang-2001-Unified}
\bea
   & & C^\alpha_\beta (k, t) = c^\alpha_\beta (k)
       - d^\alpha_\beta (k) \int^t {dt \over a^3},
   \label{GW-sol}
\eea
where the $d^\alpha_\beta$-mode also diverges logarithmically in
the same manner as $\varphi_v$ and $\varphi_\delta$. The singular
divergences occur if we reach a singularity at the bouncing epoch
with vanishing scale-factor $a$.

\section{Conclusion}

Our analyses and results are based on two important assumptions:
(i) the contracting and the expanding phases are smoothly ($a$ and
$\dot a$ are continuous) connected by a non-singular bounce, 
and (ii) the linear theory is valid.
The generated spectrum during collapsing phase with $p \simeq 0$
shows blue spectrum for $\delta \phi_\varphi$ which is identified
as the $C$-mode.  Authors of \cite{KOST-pert} find that $\delta
\phi_\chi$ has a scale-invariant spectrum, but we and \cite{BF}
have shown that it should be identified as the $d$-mode which
shows apparent growth in the contracting phase, but decays as the
model enters the expanding phase, thus uninteresting; the matching
conditions also show that $C$-mode in expanding phase is not
affected by any $d$-mode in previous history as long as the
large-scale condition is met. Although, this may sound strange
(because the growing solution in contracting phase is feeded into
the decaying one in expanding phase) it is actually apparent in
the general large-scale solutions in eqs.
(\ref{LS-sol-varphi_chi}-\ref{LS-sol-varphi_delta}).

However, as the ekpyrotic scenario encounters a singular ($a=0$)
bounce \cite{KOST,KOST-pert,KOSST}
we are not sure whether the above analyses based on
classical gravity can survive such a bounce. 
This does not mean that the other case
suggested in \cite{KOST-pert}, that through the bounce the
$d$-mode in contracting phase is switched into the $C$-mode in the
expanding phase, is plausible at all; notice that $C$ and $d$ are
coefficients of the two independent solutions. Especially, we note
that the matching conditions used in \cite{KOST-pert} are {\it ad hoc}
and {\it inconsistent} with the known matching conditions in the literature. 
\cite{Lyth-2} pointed out that as model approaches the singular epoch
the $d$-mode fluctuation grows large enough that the linear theory
could break down.
If the bouncing universe is at all possible in future
string theory context as conjectured in \cite{KOSST}, and if it
involves the singular transition, the fate of perturbations should
be handled in the context of {\it that} string theory. 
Thus, our {\it conclusion} is that either the final spectrum is blue with
suppressed amplitude or the issue should be handled in the future
string theory context with a concrete mechanism for the bounce.
In either case we find no supporting argument to accept the final
scale-invariant spectrum based on the analyses made in
\cite{KOST,KOST-pert}.

\section{Discussions}

As in the pre-big bang scenario which also gives very blue
spectra $n_S -1 \simeq 3 \simeq n_T$ \cite{Brustein-etal-1995,Hwang-1998-PBB},
in order to become a viable model to explain the large-scale structures and
the cosmic microwave background radiation anisotropy the ekpyrotic
scenario should resort to the other mechanism which is unknown at
the moment; perhaps one can find suitable parameter space in
the isocurvature modes
by considering multi-components as in the pre-big bang scenario
\cite{Copeland-etal}.
In contrast with the pre-big bang scenario where the amplitudes of the
scalar and tensor structures are comparable,
see eq. (42) in \cite{Hwang-1998-PBB}, since the scalar
structure in the ekpyrotic scenario is suppressed relative to
the tensor one, the isocurvature possibility to generate the
observed structures is more plausible, except that pure isocurvature
modes are unfavored by the large-scale structure and the cosmic microwave
background anisotropy observations \cite{isocurvature}.

Before we have the fully considered perturbations both in the
brane and the bulk, the results based on the effective field
theory should be regarded as {\it preliminary} ones. This is an
unsettled issue at the moment and whether the resulting spectra
from full consideration could be scale-invariant is far from
clear. Similar anticipation is made about whether more complete
consideration of the quantum corrections (which is actually
required as we approach the transition epoch) can make the pre-big
bang scenario a less blue and eventually scale-invariant spectra;
there is a signature in the right direction, but not enough
at the moment \cite{Cartier-etal}.

Based on the above results, we can make the following observation.
Assuming power-law expansion/contraction $a \propto |t|^p$ during
the seed generating stage from quantum fluctuations, the
observational requirement of the scale-invariant spectrum for the
scalar structure requires $p \gg 1$, thus $-1< w \ll - {1 \over
3}$. Thus, for an expanding phase we need accelerated expansion,
whereas for a contracting phase we need a damped collapse. During
the damped collapse, however, we have ${k \over a|H|}$ becoming
large as we approach the bouncing epoch $t \rightarrow -0$, thus
violating the large-scale condition we used; this can introduce a
scale dependent damping in the final spectra. 
As we have mentioned, in an undamped contraction with $p<1$ we have the
large-scale condition well met during the transition, but the resulting
spectrum is not scale-invariant.

\vskip .5cm
We thank Robert Brandenberger, Christopher Gordon, Justin Khoury,
and David Wands for useful discussions.
We also thank Paul Steinhardt and Neil Turok
for clarifying talks and explanations about their theory during
the M-theory meeting in Cambridge.
We acknowledge Robert Brandenberger for showing us the early draft
of \cite{BF} before publication,
and Andrei Linde for sharing his opinion on the issue
during COSMO-01 conference in Rovaniemi.
We wish to appreciate Neil Turok for clarifying discussions
on several aspects of the ekpyrotic scenarios and perturbations.


\end{document}